# One In-Situ Extraction Algorithm for Monitoring Bunch-by-Bunch Profile in the Storage Ring

**Ruizhe Wu, Yunkun Zhao, Leilei Tang, Jigang Wang, Ping Lu, Baogen Sun\***

National Synchrotron Radiation Laboratory (NSRL), University of Science and Technology of China (USTC), No. 96, JinZhai Road Baohe District, Hefei, Anhui 230026, China
Author to whom correspondence should be addressed: bgsun@ustc.edu.cn

**Abstract**. As the brightness of synchrotron radiation (SR) light sources improves, the operation stability of light sources is weakened. To explore various beam instability related issues in light sources, one transverse beam diagnostics system for bunch-by-bunch (BbB) profile measurement has been established at Hefei Light Source-II (HLS-II). In this paper, one in-situ extraction algorithm in the data processing backend of the system is developed for BbB profiles, so as to provide important beam information of the machine operation in time.

## 1. Introduction

The SR is the emission of relativistic electrons gyrating in magnetic field, whose spectrum covers infra-red to hard X-rays. And its features of high intensity and pulses make it one powerful tool for researching many subjects, such as condensed matter physics, materials science, biology, medicine and so on. For producing SR light, scientists have built one large scientific instrument, the SR light source. The typical layout of the SR light source is shown as figure 1.

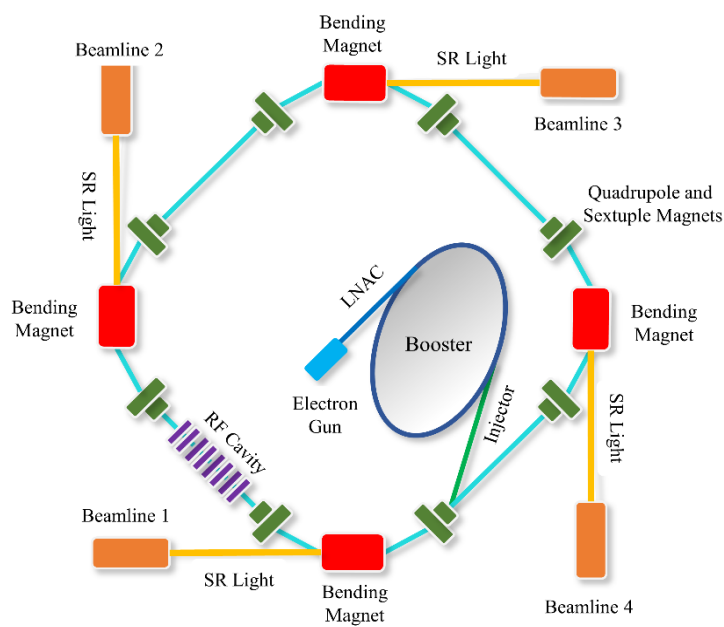

**Figure 1**. The Layout of Typical Synchrotron Radiation Light Source

Inside the light source, firstly, the beam of electrons generated from the electron gun is accelerated by the linear accelerator (LINAC) and booster to nearly the speed of light. Subsequently, the electron beam is fed into the storage ring of the light source by the injector, then bunched in the storage ring. While in storage ring, the SR light is produced when the bunched beam "bunch" passes through the bending magnets. Finally, the SR light will be guided into the beamlines, where experiments of fundamental science are conducted with the SR light. However, recently some experiments demand for producing highly brilliant beams by squeezing electron beams into extremely compact bunches in the SR light source [1]. Hence, many laboratories are building the most advanced $4^{th}$ generation SR light source, which is based on the multi-bend achromat lattice and able to surpass the brightness and coherence that are attained using present $3^{th}$ generation SR light source [2]. While with high brightness and coherence, its bunched beam gets small emittance and high current, thus resulting in deteriorations in beam stability associated with impedance concerns and coupled bunch instability [3].

To explore these instabilities, BbB technology is developed to observe the parameters of individual bunches in the storage ring on fine time scales, including bunch charge, bunch transverse positions, bunch longitudinal phase, bunch profiles and bunch length. Among them, the bunch profiles relate to the transverse quadrupole-mode beam oscillation inside the storage ring due to injection mismatch and severe quadrupole oscillation fields [4,5], which is an essential parameter of bunches for researching coupled bunch instability. In BbB technology, two approaches have been proposed to extract the bunch profiles: One is the FPGA-based approach, the other is the Oscilloscope-based approach. In the FPGA-based approach, the BbB profile can be monitored in real-time, but the overall system is too complex to build in short time [6]. While for the Oscilloscope-based approach, its system can be built quickly, and the BbB profile can be extracted by specific algorithms. However, the past extraction algorithm for BbB profile can hardly work timely due to the time cost of the Gaussian fitting process [7]. To accelerate the extraction process, one in-situ extraction algorithm for monitoring BbB profile in the storage ring was developed.

The rest of the paper is organized as follows: Section 2 overviews the Oscilloscope-based BbB profile measurement system. Section 3 details the dealing processes of the in-situ extraction algorithm. Finally, Section 4 gives a brief summary and outlook.

## 2. System Setup

The system for BbB profile measurement assembled at HLS-II light source mainly consists of two parts: an optical imaging frontend and a data processing backend, as shown in figure 2.

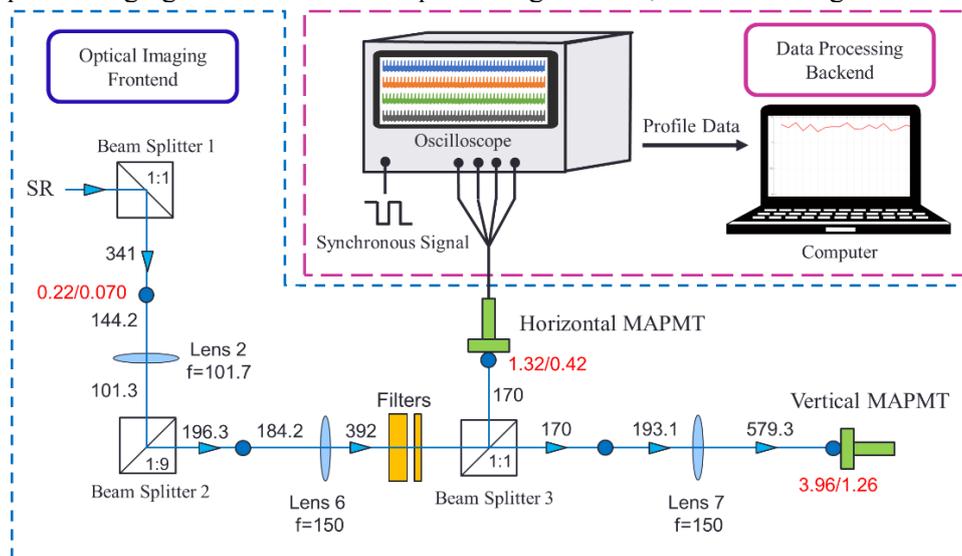

**Figure 2**. The Framework of BbB Profile Measurement System at HLS-II

In the optical imaging frontend of the system, the object of the light source point is transformed into an accurate image on the image plane of the optical platform, serving as the input for multi-anode avalanche photomultiplier tube (MAPMT) [8]. It is worth noting that the response speed of MAPMT in the frontend ranges from 4.6 to 4.9 ns, while the bunch spacing of HLS-II is 4.9 ns. Therefore, the MAPMT ensures the BbB four-channel signals do not overlap. While in the backend of data processing, the four-channel signal induced by MAPMT will be transmitted to the oscilloscope (LECORY: HDO6104A) for data acquisition. This acquisition is performed periodically, since the oscilloscope is triggered externally by the 4.534 MHz synchronous signal from the timing system of the light source. The revolution period of HLS-II is about 220 ns, and the acquisition should cover at least dozens of revolutions of profile information to better observe the profile of BbB. Hence, the sampling rate of the oscilloscope used in the backend is 10 GS/s. In its four-channel synchronous sampling mode, it can store four-channel signals with a time length of 10 μs, which contains about 45 revolutions of BbB profile information. After the acquisition, the four-channel data will be transferred to the program in the computer for extracting BbB profiles.

### 3. Algorithm Steps for Extracting BbB Profile

The algorithm consists of six dealing processes "ascertain the RF frequency of the cavity, ascertain the number of revolutions contained in acquisition, resample the four-channel data, determine the reference for BbB partition, extract BbB amplitudes of four-channel signals, calculate BbB profile information", which are detailed through the following subsections.

*3.1. Ascertain the RF frequency of the cavity*

After the four-channel signal of BbB is collected by the oscilloscope, the four-channel data will be transferred to the program of the computer for data processing. The program first performs a Fast Fourier Transformation (FFT) on the sum of four-channel data, and then the RF frequency $F_{RF}$ of the cavity in the storage ring can be found from the result of FFT, since the electron bunch is modulated by the RF frequency. As shown in figure 3, the $F_{RF}$ of the cavity is 204 MHz at HLS-II.

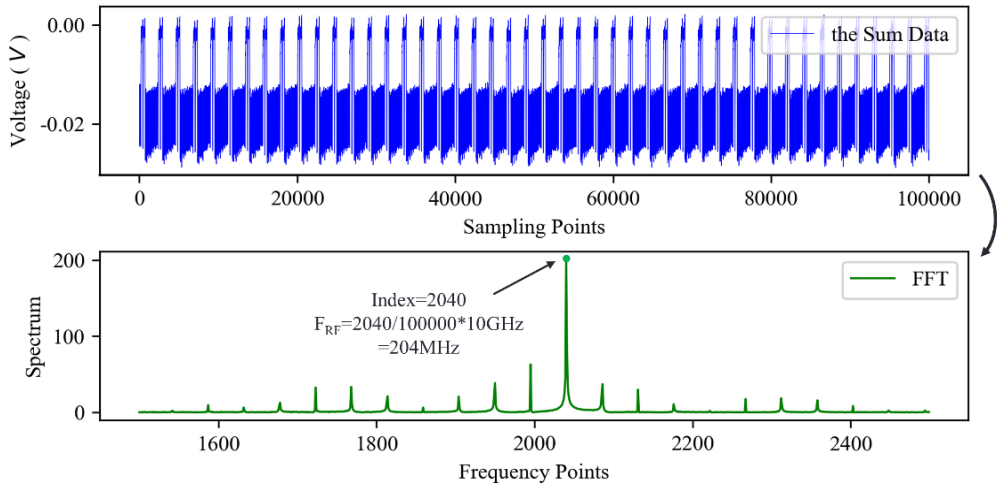

**Figure 3**. The FFT of the Sum of the Four-channel Data of the Profile

*3.2 Ascertain the number of revolutions contained in acquisition*

Using equation (2.1), the number of revolutions $N_{turns}$ in the four-channel data collected by the oscilloscope can be calculated [9]. In the backend, the four-channel data collected by the oscilloscope contains 45.34 turns.

$$N_{turns} = \frac{L \cdot F_{RF}}{h \cdot F_{sample}} \qquad (2.1)$$

Wherein: $F_{sample}$ is the sampling frequency of the oscilloscope, which is 10 GHz in the backend; $h$ is the harmonic number, which equals 45 at HLS-II; $L$ is the number of sampling data points collected by the oscilloscope, which equals 100000 in the backend.

*3.3 Resample the four-channel data*

The time length $T_{sample}$ of the four-channel data acquired by the oscilloscope is 10 μs, so the time length of the resampled data must also tend to 10 μs, from which equation (2.2) can be derived.

$$\begin{cases} Dots \cdot h \cdot N_{turns} = T_{sample} \\ T_{err} = T_{sample} - ceil(Dots) \cdot h \cdot N_{turns} \end{cases} \quad (2.2)$$

Wherein: the *ceil(\*)* indicates downward rounding; *Dots* represents the number of resampling points of a single bunch in one specific time unit; $T_{err}$ represents the cumulative time error after resampling.

Taking the number of resampling points *Dots* of one single bunch as 50, then the specific time unit of the *Dots* is 98 ps, and its corresponding cumulative time error $T_{err}$ is 2530 ps, which is about 50% of the time of bunch spacing. Since the online algorithm depends on the amplitudes of the four-channel signals within the time of bunch spacing, the cumulative time error $T_{err}$ is acceptable.

*3.4 Determine the reference for BbB partition*

Although the oscilloscope is triggered by the synchronization signal, the quantized four-channel signals still have random deviation on their time axis. Therefore, it is necessary to determine the reference for BbB partition in the resampled four-channel data.

To be specific, convolution is needed in the determination. First, one sample function with the same filling pattern as the bunches in the storage ring is constructed. Then the sample function is convolved with the first two turns of the sum of resampled four-channel data. According to the maximum absolute value of the convolution result, the corresponding moving length of the sample function can be confirmed. With this moving length, the reference for BbB partition can be determined, as shown in figure 4.

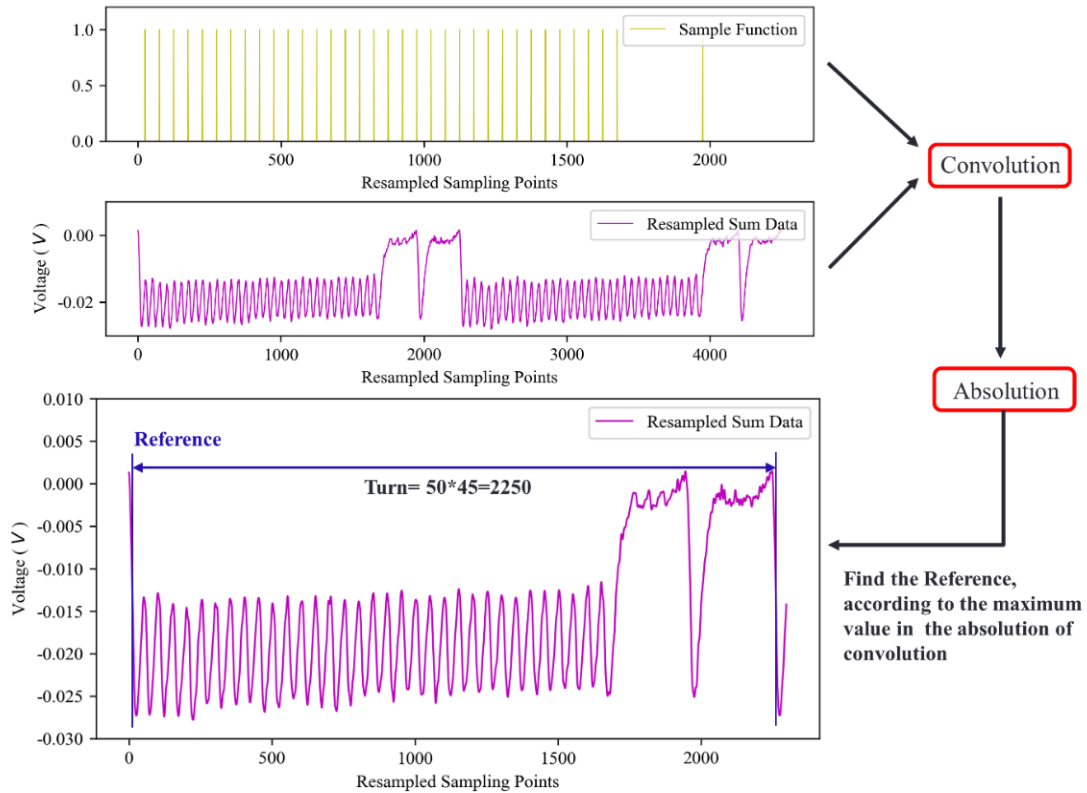

**Figure 4**. The Process of Determining the Reference of BbB Partition

*3.5 Extract BbB amplitudes of four-channel signals*

After the reference for BbB partition is determined, the individual four-channel data of each bunch can be segmented directly from the resampled four-channels data by means of index. Here, every 50 resampling data points account for the four-channel data of one bunch, and every 2250 resampling data points account for the four-channel data of one revolution of bunches.

As shown in figure 5, (a) is the sum of four-channel data of one single bunch, and (b) is the 4-channel data ranging from Channel 1 to Channel 4 of one single bunch. After finding the index of the lowest value point in the sum of four-channel data, put this index into the corresponding data of 4 channels, then the BbB amplitudes of four-channel signals can be obtained.

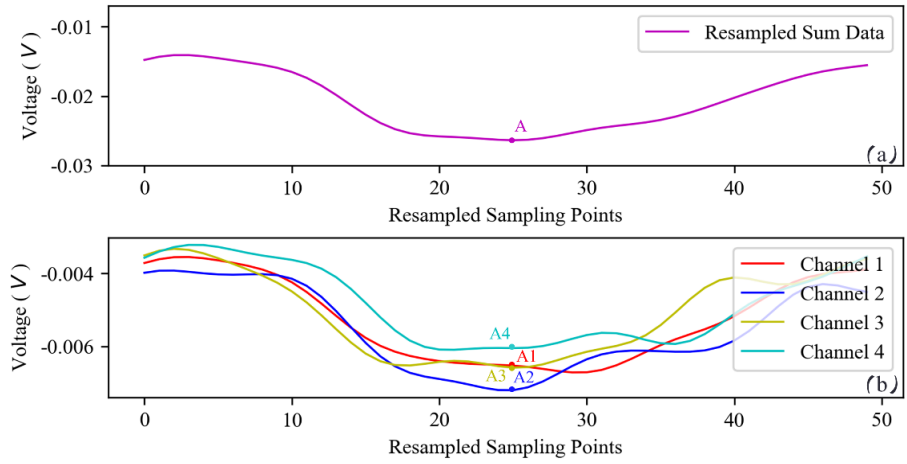

**Figure 5**. The Extracted BbB Amplitudes of Four-channels Signals

*3.6 Calculate BbB profile information*

After the BbB amplitudes A1~A4 of each bunch are extracted, the profile information of each bunch can be deduced with equation (2.3) [6]. Figure 6 shows the variation of the BbB profile of the bunch with bucket number 10 within 45 turns.

$$S_{in}(\sigma, \delta) = \left[ ln\frac{A_2 A_3}{A_1 A_4} - \tilde{d}_s \right]^{-\frac{1}{2}} \Big/ M \quad (2.3)$$

Wherein: σ denotes the size of the light spot; δ denotes the deviation of the light spot center from the reference centre at MAPMT; $\tilde{d}_s$ represents the actual gain correction factor of four channels; $M$ is the actual magnification of the optical path to the MAPMT. Among them, $M$ and $\tilde{d}_s$ are obtained with calibration.

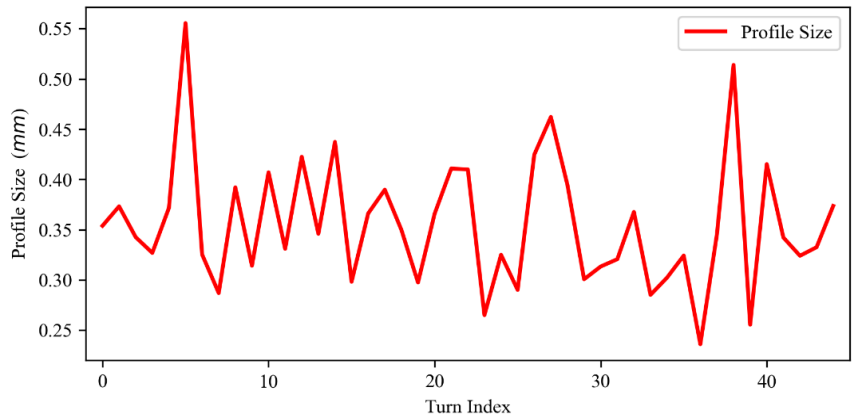

**Figure 6**. The Variation of the BbB Profile of Bunch 10 within 45 turns

## 4. Conclusion

To explore issues of beam instabilities in the light source, one BbB profile measurement system has been constructed at HLS-II. Inside the system, the algorithm for extracting BbB profiles in the data processing backend guarantees that the profile information of BbB can be timely extracted, thus providing data support for the instability analysis in the light source. In future research, we will extend the system to transverse positions of BbB and further analyze the modes of betatron oscillation in the storage ring, ensuring the stability of machine performance.